\documentclass[10pt]{article}

\usepackage[preprint]{tmlr}

\usepackage{hyperref}
\usepackage{url}
\usepackage{booktabs}
\usepackage{amsmath}
\usepackage{array}
\usepackage{graphicx}

\title{RAGFlip: Measuring Query-Level Negative Flips in Retriever Upgrades}

\author{
\name Elyas Irankhah \email Elyas.irankhah@yale.edu \\
\addr Department of Emergency Medicine, Yale School of Medicine
\AND
\name Muhammad Arif \email muhammad.arif@yale.edu \\
\addr Department of Emergency Medicine, Yale School of Medicine
}

\def\month{October}
\def\year{2026}
\def\openreview{}

\begin{document}

\maketitle

\begin{abstract}
Retriever upgrades are typically evaluated using aggregate metrics, which can hide regressions on queries the previous retriever already served correctly.
We study these regressions as \emph{negative flips}: queries for which BM25 retrieves a judged relevant passage and the replacement does not.
We evaluate BGE-large, E5-large-v2, and SPLADE on three BEIR collections Natural Questions, HotpotQA, and FiQA across five retrieval depths.
All replacements improve overall retrieval coverage. Negative flips occur in every setting and vary substantially by corpus, retriever, and depth.
At $k{=}1$, 8.6--37.5\% of BM25 successes are lost across the evaluated settings. Negative-flip rates are lower in the larger-depth settings, where the BM25-supported cohort is defined separately at each depth. These rates use the any-relevant support label. On HotpotQA at $k{=}10$, requiring every positive qrel passage raises the negative-flip rate to 12.3--17.3\%.
Simple fixed-budget combinations with BM25 reduce these regressions, and a HotpotQA reader experiment provides a limited downstream check in which some retrieval flips are accompanied by answer regressions.
These results motivate evaluating retriever updates using query-level compatibility alongside aggregate retrieval quality.
\end{abstract}

\section{Introduction}

Retrievers are commonly evaluated using aggregate ranking metrics \citep{craswell2020trec}. In open-domain question answering, a common retrieval criterion is whether relevant evidence appears within the top $k$ results \citep{karpukhin2020dense,thakur2021beir}. Modern dense retrievers can substantially improve aggregate retrieval quality over traditional lexical baselines in many settings \citep{karpukhin2020dense,wang2022e5}. However, aggregate improvements can hide losses on individual queries \citep{wang2012robust}. A replacement retriever may therefore recover queries that the existing system misses while losing previously successful ones. This concern is closely related to negative flips in model updates. \citet{yan2021positive} define a negative flip as an example that is predicted correctly by an old model and incorrectly by its replacement, demonstrating that aggregate performance and backward-compatible behavior capture different properties of an update. Backward compatibility has subsequently been studied in NLP model upgrades and first-stage retrieval \citep{lai2023improving,cai2023l2r}. Our focus is \emph{behavioral compatibility}: whether queries supported by an existing retriever remain supported after replacement. We introduce \textsc{RAGFlip}, which characterizes query-level transitions as preserved, fixed, broken, or unchanged and summarizes them using fix rate, negative-flip rate, compatibility, and net gain. Using BM25 as the reference we evaluate BGE-large-en-v1.5, E5-large-v2, and SPLADE. Because evidence identification and answer generation represent distinct stages of question answering \citep{irankhah2026archehrqa}, we also conduct a secondary HotpotQA experiment examining whether retrieval negative flips are accompanied by answer regressions for a frozen 7B reader. We organize the study around four questions:

\begin{enumerate}
    \item Can improvements in aggregate retrieval quality coexist with regressions on queries that BM25 previously served correctly?
    \item How does the negative-flip rate vary with retrieval depth, corpus, and replacement retriever?
    \item At a fixed passage budget, to what extent can standard hybrid rankings reduce these regressions?
    \item How do negative-flip rates vary with the rank at which BM25 retrieved its first relevant passage?
\end{enumerate}

To answer these questions, we conduct a systematic empirical analysis of behavioral compatibility in retriever upgrades. Specifically, we:
(i) formulate retriever replacement as a query-level transition problem that distinguishes preserved, fixed, and broken queries;
(ii) measure these transitions across three retrievers, three corpora, and five retrieval depths, with uncertainty estimates and paired comparisons;
(iii) evaluate fixed-budget hybrid rankings as simple strategies for reducing negative flips; and
(iv) examine, through a limited HotpotQA reader experiment, whether retrieval-level regressions are accompanied by answer regressions.

\section{Related Work}

\paragraph{Compatibility and regression in model updates.}
Backward compatibility has been studied as a property of evolving machine-learning systems \citep{srivastava2020backward}. In representation learning, \citet{shen2020backward} train new embedding models so that new representations remain comparable with those produced by an older model. This avoids recomputing an existing retrieval gallery. More recently, \citet{schumann2024backward} study prediction regressions caused by model updates and propose weight interpolation to reduce negative flips while preserving gains from the updated model. This line of work treats compatibility as an explicit objective during training or model updating. Our setting is different. We do not modify or retrain the replacement retriever. We evaluate the query-level regressions that arise when an independently trained retriever replaces an existing system. Related work in risk-sensitive information retrieval has shown that gains in average ranking effectiveness can coexist with losses on individual queries \citep{wang2012robust}. RAGFlip extends this perspective to retriever replacement by directly measuring before-and-after query transitions.

\paragraph{Neural retrieval architectures and matching signals.}
Neural retrieval systems differ in how they represent and match queries and documents. ANCE (Approximate Nearest Neighbor Negative Contrastive Estimation) trains a dense retriever using hard negatives selected from an approximate nearest-neighbor index of the corpus \citep{xiong2021ance}. ColBERT (Contextualized Late Interaction over BERT) represents queries and documents with token-level embeddings and computes relevance through late interaction between these representations \citep{khattab2020colbert}. COIL (COntextualized Inverted List) combines exact lexical matching with contextualized token representations stored in inverted lists \citep{gao2021coil}. These architectures use different matching mechanisms and can therefore behave differently across queries. This motivates evaluating retrieval changes at the query level in addition to aggregate effectiveness.

\paragraph{Combining lexical and neural retrieval.}
Lexical and neural retrieval signals often capture different relevance cues. \citet{wang2021interpolation} show that interpolating BM25 with BERT-based dense retrievers can substantially improve passage retrieval, with part of the gain attributed to lexical matching recovering signals missed by dense retrieval. Rank-fusion methods provide a model-independent way to combine retrieval systems. Reciprocal rank fusion combines rankings using only rank positions and has long been used as a robust aggregation method \citep{cormack2009rrf}. In our experiments, these combinations serve as fixed-budget baselines for testing whether retaining evidence from BM25 can reduce regressions introduced by its replacement.

\section{Experimental Setup}
\label{sec:setup}

\subsection{Corpora and retrievers}

Experiments are conducted on three BEIR test collections \citep{thakur2021beir}: Natural Questions \citep{kwiatkowski2019natural}, HotpotQA \citep{yang2018hotpotqa}, and FiQA \citep{maia2018fiqa}, containing 3,452, 7,405, and 648 test queries, respectively. Table~\ref{tab:corpora} summarizes the three BEIR test collections used in the study. Retrieval is performed over each full collection, with passages formed by concatenating title and text. A query is supported at depth $k$ when at least one passage with a positive BEIR qrel appears in the top-$k$ results. The reported HotpotQA results do not require every positive qrel passage to be retrieved. BM25, with $k_1{=}1.5$ and $b{=}0.75$ \citep{robertson2009probabilistic}, is the reference retriever. The replacements are \texttt{BAAI/bge-large-en-v1.5} \citep{xiao2023cpack}, \texttt{intfloat/e5-large-v2} \citep{wang2022e5}, and \texttt{naver/splade-cocondenser-ensembledistil} \citep{formal2021splade,formal2022splade}. For every model, the top-50 ranking is computed once, and smaller depths are truncations of that ranking. No retriever is tuned on the test queries. Primary labels use only BEIR relevance judgments. BM25 lowercases text, replaces non-alphanumeric characters with spaces, and removes the same 43-word list from queries and documents on every corpus: \emph{the, a, an, of, in, to, and, is, are, was, were, for, on, that, this, what, which, who, whom, whose, when, where, why, how, did, does, do, with, from, by, or, as, at, be, been, it, its, their, his, her, they, them, there}. Document frequency uses the same tokenization. The inverse document frequency weight is $\log\bigl(1+(N-\mathrm{df}+0.5)/(\mathrm{df}+0.5)\bigr)$. BGE queries are prefixed with ``Represent this sentence for searching relevant passages: ''. Passages are encoded as the title and text joined by ``. '', with no passage prefix. E5 queries use the prefix ``query: '' and passages use ``passage: ''. Both dense models are encoded with \texttt{SentenceTransformer.encode}, L2-normalized, and compared by inner product. Pooling and tokenization are those of the checkpoint; the scripts do not add a separate pooling step. The E5 runs and the FiQA BGE run set the maximum sequence length to 512. The Natural Questions and HotpotQA BGE runs leave the checkpoint default unchanged. SPLADE uses the checkpoint tokenizer and masked-language-model head. Inputs are truncated to 256 tokens. Token scores are $\log(1+\mathrm{ReLU}(\cdot))$, max-pooled over positions, and restricted to the 256 largest positive weights for both queries and passages. Ranking uses the resulting sparse dot product. The 256-token limit is the input setting of this evaluation. The 256-weight cutoff is an additional sparsification step. An implementation of RAGFlip, including the command-line checker and reproduction scripts, is available at \url{https://github.com/Elyasirankhah/RAGFlip}.

\begin{table}[t]
\caption{BEIR test collections used in the study.}
\label{tab:corpora}
\centering
\begin{tabular}{lrrl}
\toprule
Corpus & Passages & Questions & Domain \\
\midrule
Natural Questions & 2,681,468 & 3,452 & Wikipedia, single-hop QA \\
HotpotQA & 5,233,329 & 7,405 & Wikipedia, multi-hop QA \\
FiQA & 57,638 & 648 & Financial QA \\
\bottomrule
\end{tabular}
\end{table}

\subsection{Transition metrics}

To evaluate retriever replacement at the query level, we track how each query changes when BM25 is replaced by a new retriever. This formulation separates newly recovered queries from previously successful queries that become unsupported, in addition to reporting aggregate support. For query $i$, let

\[
B_i =
\begin{cases}
1, & \text{if BM25 is supported at depth } k,\\
0, & \text{otherwise},
\end{cases}
\qquad
D_i =
\begin{cases}
1, & \text{if the replacement retriever is supported at depth } k,\\
0, & \text{otherwise}.
\end{cases}
\]

Each query therefore falls into one of four transition states:

\begin{center}
\begin{tabular}{lcc}
\toprule
& $D_i=1$ & $D_i=0$ \\
\midrule
$B_i=1$ & Preserved ($C$) & Negative flip ($R$) \\
$B_i=0$ & Fixed ($F$) & Unchanged failure \\
\bottomrule
\end{tabular}
\end{center}

A preserved query is supported by both retrievers. A negative flip is supported by BM25 and unsupported by the replacement. A fixed query is unsupported by BM25 and recovered by the replacement. Their counts are:

\begin{align}
C &= \sum_{i=1}^{n} B_iD_i, \\
R &= \sum_{i=1}^{n} B_i(1-D_i), \\
F &= \sum_{i=1}^{n} (1-B_i)D_i.
\end{align}

Let $n_B=\sum_i B_i$ denote the number of BM25-supported queries and $n_{\neg B}=n-n_B$ the number of BM25 failures. We summarize replacement behavior using four metrics:

\begin{align}
\text{FixRate} &= \frac{F}{n_{\neg B}}, \\
\text{NegativeFlipRate} &= \frac{R}{n_B}, \\
\text{Compatibility} &= \frac{C}{n_B}
                      = 1-\frac{R}{n_B}, \\
\text{NetGain} &= F-R.
\end{align}

Let $X_i=+1$ for a fixed query, $X_i=-1$ for a negative flip, and $X_i=0$ otherwise, so $\mathrm{NetGain}=\sum_i X_i$. With $p_F=F/n$ and $p_R=R/n$, the plug-in variance of this sum is
\begin{equation}
\mathrm{Var}(\mathrm{NetGain})
=
n\bigl[p_F+p_R-(p_F-p_R)^2\bigr].
\end{equation}
The reported interval is the 95\% Wald interval $\mathrm{NetGain}\pm 1.96\sqrt{\mathrm{Var}(\mathrm{NetGain})}$. Fixed and broken queries are paired outcomes on the same $n$ questions, so $F$ and $R$ are not treated as independent samples. FixRate is the fraction of BM25 failures recovered by the replacement. NegativeFlipRate is the fraction of BM25 successes that are lost. Compatibility is the fraction of BM25 successes that remain supported. A positive net gain can occur together with regressions on previously successful queries. We report 95\% Wilson confidence intervals for the rates \citep{wilson1927probable}, as shown in Tables~\ref{tab:depth}, \ref{tab:depth-splade}, \ref{tab:k10rates}, and \ref{tab:rank}. BGE and E5 are compared with a two-sided exact McNemar test on BM25-supported queries \citep{mcnemar1947note}. Within each corpus, the five depth-specific tests are adjusted with Holm's method \citep{holm1979simple}. 

\subsection{Fixed-budget mixes and union baseline}

We evaluate two fixed-budget combinations of BM25 and each replacement retriever: rank alternation and reciprocal rank fusion (RRF). Each method returns exactly $k$ distinct passages. Rank alternation interleaves the two ranked lists. It selects one passage from the replacement retriever, followed by one from BM25. Duplicate passages are skipped until $k$ distinct passages are collected. RRF combines the two rankings using reciprocal rank scores \citep{cormack2009rrf}. For passage $d$:

\[
\mathrm{RRF}(d)
=
\sum_{r:\,d\in r}
\frac{1}{60+\operatorname{rank}_{r}(d)},
\]

where $r$ denotes either the BM25 ranking or the replacement ranking. The $k$ passages with the highest RRF scores are retained. We also evaluate the union of the BM25 and replacement top-$k$ lists. The union can contain up to $2k$ passages. It retains every BM25 top-$k$ passage, so it has zero qrel-defined negative flips by construction. We use this setting as a cost baseline because the passage budget can double.

\subsection{Protocol}

We evaluate five pre-specified retrieval depths, $k\in\{1,5,10,20,50\}$. The same relevance definition, BM25 parameters, tokenization, and stopword list are used on every corpus. Dense embeddings, SPLADE weights, and top-50 rankings are computed once and checkpointed. Smaller depths truncate those rankings, so every depth uses the same retrieval output. The reader experiment is a secondary check on one dataset. Recomputing the reference BM25 ranking with this tokenization and these parameters reproduced the saved top-50 list on every query of all three corpora. Recomputing the primary HotpotQA counts from the saved rankings reproduced the reported counts at all five depths.

\subsection{Downstream answer check}
\label{sec:answers}

We examine whether retrieval negative flips on HotpotQA at $k{=}10$ are accompanied by answer regressions. For each retriever, the analysis includes all negative-flip queries and a same-size control sample drawn from queries supported by both BM25 and the replacement. The resulting flip sets contain 143 BGE, 156 E5, and 125 SPLADE queries. Control queries are sampled using a fixed random seed. A frozen \texttt{Qwen/Qwen2.5-7B-Instruct} reader \citep{qwen2025qwen25} receives each question together with the ten retrieved passages. Each passage is truncated to 300 words. Generation uses greedy decoding with a maximum of 24 new tokens. The model is instructed to provide a short answer based on the retrieved passages or return \emph{unknown}. Predictions and gold answers are lowercased and stripped of punctuation. The articles \emph{a}, \emph{an}, and \emph{the} are also removed. A prediction is considered correct when the normalized gold-token sequence appears as a contiguous subsequence of the normalized prediction. The output \emph{unknown} is counted as incorrect. An answer regression occurs when the BM25 passages produce a correct answer and the replacement passages produce an incorrect answer. Natural Questions is excluded because the available short-answer resource matches fewer than 90\% of its BEIR questions, and FiQA is excluded because the analysis requires short reference answers.

\section{Results}

\subsection{Aggregate gains coexist with negative flips}

\begin{figure}[t]
\centering
\includegraphics[width=\linewidth]{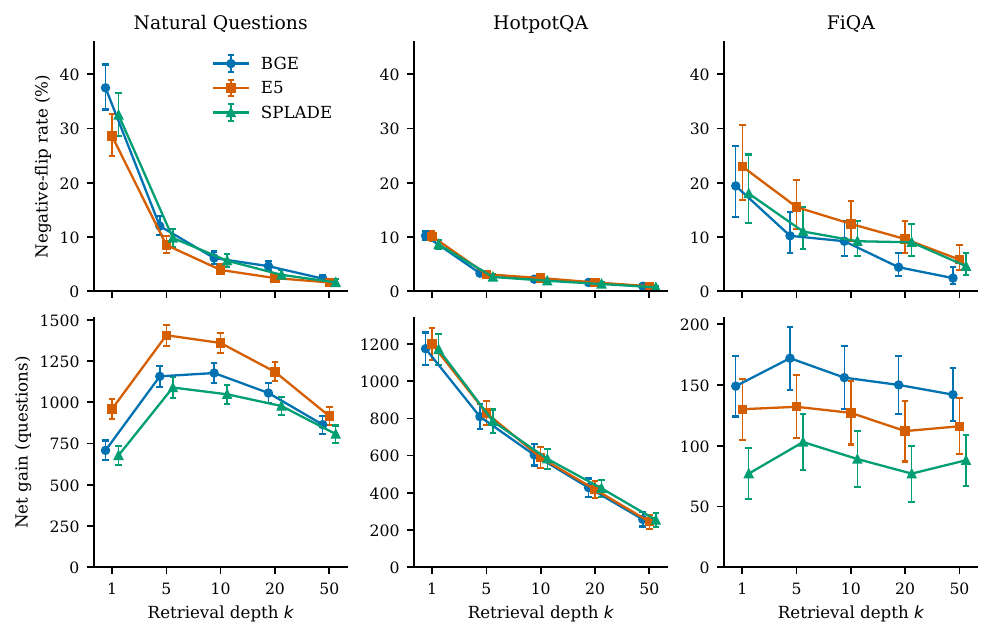}
\caption{Negative-flip rate and net gain across five retrieval depths. Bars show 95\% intervals. Flip-rate panels share a vertical scale; net-gain panels do not because corpus sizes differ.}
\label{fig:depth}
\end{figure}

Table~\ref{tab:transitions} reports the query-level transitions at $k{=}10$.
Every replacement supports more queries overall than BM25, and every comparison also contains negative flips. For example, E5 increases Natural Questions support from 1,600 to 2,959 queries by fixing 1,422 BM25 failures and losing 63 BM25 successes. On FiQA, BGE fixes 184 failures and breaks 28 previous successes. Figure~\ref{fig:depth} and Tables~\ref{tab:depth}--\ref{tab:depth-splade} show these transitions across retrieval depth. In every retriever--corpus series, the negative-flip rate is highest in the $k{=}1$ setting and lowest in the $k{=}50$ setting, and net gain is positive at every depth. The 95\% intervals for the flip rates at $k{=}1$ and $k{=}50$ do not overlap in any series. The magnitude depends strongly on the corpus. At $k{=}1$, negative-flip rates for the two dense retrievers are 28.6--37.5\% on Natural Questions, 10.1--10.2\% on HotpotQA, and 19.4--23.0\% on FiQA. At $k{=}10$, the three replacements have rates of 3.9--6.1\% on Natural Questions, 1.9--2.4\% on HotpotQA, and 9.2--12.4\% on FiQA. Table~\ref{tab:k10rates} separates recovery from preservation at $k{=}10$. Compatibility exceeds 93\% on Natural Questions and 97\% on HotpotQA for all three replacements, and is lower on FiQA. Fix rates also vary by corpus: E5 recovers 76.8\% of BM25 failures on Natural Questions and 48.2\% on FiQA. Comparisons across $k$ are comparisons across deployment settings. The BM25-supported cohort is defined separately at each depth, so these curves are not a fixed-cohort comparison. A within-BM25 comparison holds the tokenizer, stop list, and qrels fixed and changes one parameter at a time. BM25-$k_1$ uses $k_1{=}1.2$ and $b{=}0.75$. BM25-$b$ uses $k_1{=}1.5$ and $b{=}0.65$. The reference, $k_1{=}1.5$ and $b{=}0.75$, matches the saved top-50 ranking on every query of all three corpora. At $k{=}10$, the two variants break 20 and 23 of the 1,600 Natural Questions successes, 9 and 9 of the 6,484 HotpotQA successes, and 2 and 4 of the 306 FiQA successes. On those same cohorts the three replacements break 63--97, 125--156, and 28--38 successes. At every corpus and every evaluated depth, each parameter variant breaks fewer reference successes than BGE, E5, or the evaluated SPLADE configuration. The variants also recover some reference misses. HotpotQA is also evaluated under a strict support label: every positive qrel passage must appear in the top $k$. Recomputing the primary any-relevant counts from the saved rankings reproduces the reported HotpotQA transition counts at all five depths. Under the strict label, BM25 supports 0 queries at $k{=}1$, 1,874 at $k{=}5$, 2,432 at $k{=}10$, 2,974 at $k{=}20$, and 3,644 at $k{=}50$. The strict flip rate is undefined at $k{=}1$ because that BM25-supported denominator is zero. At $k{=}10$, BGE, E5, and SPLADE break 334, 298, and 420 of the 2,432 strict successes (13.7\%, 12.3\%, and 17.3\%). Under the any-relevant label those rates are 2.2\%, 2.4\%, and 1.9\% of 6,484 successes. For each replacement, the strict negative-flip rate is lower at $k{=}50$ than at $k{=}5$, and net gain is positive at every depth with a nonzero denominator. The main tables keep the any-relevant label.

\begin{table}[t]
\caption{Query-level transitions at $k{=}10$. ``Both'' denotes queries supported by both BM25 and the replacement.}
\label{tab:transitions}
\centering
\small
\begin{tabular}{llrrrrr}
\toprule
Corpus & Model & BM25 & New & Both & Fixed & Broken \\
\midrule
Natural Questions & BGE & 1600 & 2777 & 1503 & 1274 & 97 \\
Natural Questions & E5 & 1600 & 2959 & 1537 & 1422 & 63 \\
Natural Questions & SPLADE & 1600 & 2648 & 1511 & 1137 & 89 \\
HotpotQA & BGE & 6484 & 7088 & 6341 & 747 & 143 \\
HotpotQA & E5 & 6484 & 7073 & 6328 & 745 & 156 \\
HotpotQA & SPLADE & 6484 & 7067 & 6359 & 708 & 125 \\
FiQA & BGE & 306 & 462 & 278 & 184 & 28 \\
FiQA & E5 & 306 & 433 & 268 & 165 & 38 \\
FiQA & SPLADE & 306 & 395 & 278 & 117 & 28 \\
\bottomrule
\end{tabular}
\end{table}

\begin{table}[t]
\caption{Negative-flip rate and net gain for the dense retrievers. Brackets show 95\% intervals.}
\label{tab:depth}
\centering
\small
\begin{tabular}{lcccc}
\toprule
& \multicolumn{2}{c}{BGE} & \multicolumn{2}{c}{E5} \\
\cmidrule(lr){2-3} \cmidrule(lr){4-5}
$k$ & Flip (\%) & Net & Flip (\%) & Net \\
\midrule
\multicolumn{5}{l}{\emph{Natural Questions, 3,452 questions}} \\
1  & 37.5 [33.5, 41.8] & +708 [648, 768] & 28.6 [24.9, 32.7] & +959 [897, 1021] \\
5  & 12.0 [10.3, 13.9] & +1157 [1093, 1221] & 8.5 [7.0, 10.2] & +1406 [1343, 1469] \\
10 & 6.1 [5.0, 7.3] & +1177 [1116, 1238] & 3.9 [3.1, 5.0] & +1359 [1299, 1419] \\
20 & 4.6 [3.7, 5.6] & +1056 [997, 1115] & 2.4 [1.8, 3.2] & +1184 [1126, 1242] \\
50 & 2.3 [1.8, 3.0] & +864 [810, 918] & 1.6 [1.1, 2.2] & +916 [862, 970] \\
\midrule
\multicolumn{5}{l}{\emph{HotpotQA, 7,405 questions}} \\
1  & 10.2 [9.4, 11.1] & +1175 [1088, 1262] & 10.1 [9.3, 11.0] & +1200 [1113, 1287] \\
5  & 3.3 [2.9, 3.8] & +810 [744, 876] & 3.1 [2.7, 3.5] & +829 [764, 894] \\
10 & 2.2 [1.9, 2.6] & +604 [547, 661] & 2.4 [2.1, 2.8] & +589 [532, 646] \\
20 & 1.6 [1.3, 1.9] & +428 [379, 477] & 1.6 [1.3, 1.9] & +418 [370, 466] \\
50 & 0.9 [0.7, 1.1] & +257 [219, 295] & 0.9 [0.7, 1.2] & +245 [207, 283] \\
\midrule
\multicolumn{5}{l}{\emph{FiQA, 648 questions}} \\
1  & 19.4 [13.7, 26.8] & +149 [124, 174] & 23.0 [16.8, 30.7] & +130 [105, 155] \\
5  & 10.2 [7.0, 14.6] & +172 [146, 198] & 15.5 [11.5, 20.5] & +132 [106, 158] \\
10 & 9.2 [6.4, 12.9] & +156 [130, 182] & 12.4 [9.2, 16.6] & +127 [101, 153] \\
20 & 4.4 [2.7, 7.0] & +150 [126, 174] & 9.6 [7.0, 13.0] & +112 [87, 137] \\
50 & 2.4 [1.3, 4.4] & +142 [120, 164] & 5.8 [3.9, 8.5] & +116 [93, 139] \\
\bottomrule
\end{tabular}
\end{table}

\begin{table}[t]
\caption{SPLADE negative-flip rate and net gain. Brackets show 95\% intervals.}
\label{tab:depth-splade}
\centering
\small
\begin{tabular}{lcc}
\toprule
$k$ & Flip (\%) & Net \\
\midrule
\multicolumn{3}{l}{\emph{Natural Questions, 3,452 questions}} \\
1  & 32.5 [28.6, 36.6] & +678 [620, 736] \\
5  & 9.8 [8.2, 11.5] & +1089 [1027, 1151] \\
10 & 5.6 [4.5, 6.8] & +1048 [989, 1107] \\
20 & 3.0 [2.3, 3.9] & +977 [921, 1033] \\
50 & 1.6 [1.2, 2.2] & +807 [755, 859] \\
\midrule
\multicolumn{3}{l}{\emph{HotpotQA, 7,405 questions}} \\
1  & 8.6 [7.8, 9.4] & +1171 [1087, 1255] \\
5  & 2.6 [2.3, 3.1] & +785 [722, 848] \\
10 & 1.9 [1.6, 2.3] & +583 [528, 638] \\
20 & 1.3 [1.0, 1.6] & +424 [377, 471] \\
50 & 0.7 [0.6, 1.0] & +254 [217, 291] \\
\midrule
\multicolumn{3}{l}{\emph{FiQA, 648 questions}} \\
1  & 18.0 [12.5, 25.2] & +77 [56, 98] \\
5  & 11.0 [7.7, 15.5] & +103 [80, 126] \\
10 & 9.2 [6.4, 12.9] & +89 [66, 112] \\
20 & 9.0 [6.5, 12.4] & +77 [54, 100] \\
50 & 4.6 [3.0, 7.1] & +88 [67, 109] \\
\bottomrule
\end{tabular}
\end{table}

\begin{table}[t]
\caption{Fix rate and compatibility at $k{=}10$. Brackets show 95\% Wilson intervals.}
\label{tab:k10rates}
\centering
\small
\begin{tabular}{llcc}
\toprule
Corpus & Model & Fix rate (\%) & Compatibility (\%) \\
\midrule
Natural Questions & BGE & 68.8 [66.6, 70.9] & 93.9 [92.7, 95.0] \\
Natural Questions & E5 & 76.8 [74.8, 78.7] & 96.1 [95.0, 96.9] \\
Natural Questions & SPLADE & 61.4 [59.2, 63.6] & 94.4 [93.2, 95.5] \\
HotpotQA & BGE & 81.1 [78.5, 83.5] & 97.8 [97.4, 98.1] \\
HotpotQA & E5 & 80.9 [78.2, 83.3] & 97.6 [97.2, 97.9] \\
HotpotQA & SPLADE & 76.9 [74.0, 79.5] & 98.1 [97.7, 98.4] \\
FiQA & BGE & 53.8 [48.5, 59.0] & 90.8 [87.1, 93.6] \\
FiQA & E5 & 48.2 [43.0, 53.5] & 87.6 [83.4, 90.8] \\
FiQA & SPLADE & 34.2 [29.4, 39.4] & 90.8 [87.1, 93.6] \\
\bottomrule
\end{tabular}
\end{table}

\subsection{Fixed-budget combinations reduce negative flips}

Table~\ref{tab:mix} compares each replacement alone with rank alternation and RRF at a fixed budget of ten passages. Both combinations reduce the number of broken BM25 successes. On Natural Questions, that reduction comes with lower net gain. For BGE, RRF reduces the broken count from 97 to 23 and net gain from $+1{,}177$ to $+873$. For E5, the corresponding change is from 63 to 12 broken queries and from $+1{,}359$ to $+997$ net gain. On HotpotQA, rank alternation reduces broken queries and increases net gain for all three replacements. RRF improves both quantities for BGE and E5. For SPLADE, RRF reduces the broken count from 125 to 41 while net gain decreases from $+583$ to $+542$. On FiQA, both combinations reduce the broken count, and net gain is lower than the replacement alone in every row. The union baseline has zero retrieval negative flips by construction because every BM25 top-10 passage is retained. At $k{=}10$, the union uses a context of up to 20 passages. The corresponding net gains are $+1{,}274$, $+1{,}422$, and $+1{,}137$ on Natural Questions; $+747$, $+745$, and $+708$ on HotpotQA; and $+184$, $+165$, and $+117$ on FiQA for BGE, E5, and SPLADE, respectively.

\begin{table}[t]
\caption{Broken queries and net gain at a fixed budget of ten passages.}
\label{tab:mix}
\centering
\small
\begin{tabular}{llrrrrrr}
\toprule
& & \multicolumn{2}{c}{Alone} & \multicolumn{2}{c}{Alternate} & \multicolumn{2}{c}{RRF} \\
\cmidrule(lr){3-4} \cmidrule(lr){5-6} \cmidrule(lr){7-8}
Corpus & Model & Broken & Net & Broken & Net & Broken & Net \\
\midrule
NQ & BGE & 97 & +1177 & 48 & +1024 & 23 & +873 \\
NQ & E5 & 63 & +1359 & 32 & +1205 & 12 & +997 \\
NQ & SPLADE & 89 & +1048 & 40 & +937 & 13 & +726 \\
HotpotQA & BGE & 143 & +604 & 40 & +669 & 24 & +677 \\
HotpotQA & E5 & 156 & +589 & 37 & +671 & 31 & +645 \\
HotpotQA & SPLADE & 125 & +583 & 45 & +615 & 41 & +542 \\
FiQA & BGE & 28 & +156 & 17 & +143 & 6 & +137 \\
FiQA & E5 & 38 & +127 & 20 & +115 & 15 & +120 \\
FiQA & SPLADE & 28 & +89 & 15 & +84 & 8 & +69 \\
\bottomrule
\end{tabular}
\end{table}

\subsection{Negative-Flip Rates Vary with BM25 Rank}

Table~\ref{tab:rank} is a descriptive stratification of $k{=}10$ qrel-defined negative flips by the rank of BM25's first relevant passage, for BGE and E5. The lowest point estimate is the rank-1 stratum in every model--corpus pair. The difference between rank 1 and ranks 6--10 is larger on HotpotQA and FiQA than on Natural Questions. 

\begin{table}[t]
\caption{Descriptive negative flips at $k{=}10$ by the rank of BM25's first relevant passage. Each cell is flips/denominator (rate in percent [95\% Wilson interval]).}
\label{tab:rank}
\centering
\footnotesize
\begin{tabular}{llccc}
\toprule
Corpus & Model & Rank 1 & Ranks 2--5 & Ranks 6--10 \\
\midrule
NQ & BGE & 24/517 (4.6 [3.1, 6.8]) & 52/723 (7.2 [5.5, 9.3]) & 21/360 (5.8 [3.9, 8.8]) \\
NQ & E5 & 19/517 (3.7 [2.4, 5.7]) & 28/723 (3.9 [2.7, 5.5]) & 16/360 (4.4 [2.8, 7.1]) \\
HotpotQA & BGE & 74/4875 (1.5 [1.2, 1.9]) & 42/1255 (3.4 [2.5, 4.5]) & 27/354 (7.6 [5.3, 10.9]) \\
HotpotQA & E5 & 79/4875 (1.6 [1.3, 2.0]) & 50/1255 (4.0 [3.0, 5.2]) & 27/354 (7.6 [5.3, 10.9]) \\
FiQA & BGE & 5/139 (3.6 [1.6, 8.1]) & 11/107 (10.3 [5.8, 17.5]) & 12/60 (20.0 [11.8, 31.8]) \\
FiQA & E5 & 6/139 (4.3 [2.0, 9.1]) & 17/107 (15.9 [10.2, 24.0]) & 15/60 (25.0 [15.8, 37.2]) \\
\bottomrule
\end{tabular}
\end{table}

\subsection{Relative compatibility is corpus dependent}

BGE and E5 do not have a consistent ordering in negative-flip behavior.
Table~\ref{tab:paired} reports their paired disagreements at $k{=}10$. On Natural Questions, BGE loses more BM25-supported queries than E5 at every tested depth. At $k{=}10$, 60 queries flip only under BGE and 26 only under E5 ($p{=}0.0003$, Holm-adjusted $p{=}0.0006$). The adjusted difference remains significant at all five depths. On HotpotQA, the two dense retrievers are not detectably different at any depth. On FiQA, the direction reverses: E5 produces more unique flips, with significant adjusted differences at $k{=}20$ and $k{=}50$. No dense replacement is uniformly more compatible with BM25 across the three corpora.

\begin{table}[t]
\caption{Paired negative flips at $k{=}10$ among BM25-supported queries. ``Only BGE'' denotes queries lost by BGE and preserved by E5.}
\label{tab:paired}
\centering
\begin{tabular}{lrrrr}
\toprule
Corpus & Only BGE & Only E5 & $p$ & $p_{\mathrm{Holm}}$ \\
\midrule
Natural Questions & 60 & 26 & 0.0003 & 0.0006 \\
HotpotQA & 64 & 77 & 0.31 & 1.00 \\
FiQA & 10 & 20 & 0.099 & 0.197 \\
\bottomrule
\end{tabular}
\end{table}

\subsection{Retrieval flips do not always become answer regressions}
\label{sec:answer-results}

Table~\ref{tab:answers} reports the HotpotQA reader analysis at $k{=}10$.
Among retrieval negative flips, answer regressions occur for 24 of 143 BGE queries (16.8\%), 33 of 156 E5 queries (21.2\%), and 19 of 125 SPLADE queries (15.2\%). The corresponding rates in the same-size control samples are 6.3\%, 1.3\%, and 7.2\%. Most retrieval flips occur on questions that the reader already answers incorrectly from the BM25 passages: 119 of 143 BGE queries (83.2\%), 121 of 156 E5 queries (77.6\%), and 105 of 125 SPLADE queries (84.0\%). Among the smaller subset of flip queries answered correctly from BM25 passages, the replacement answer becomes incorrect in 24/24 BGE cases, 33/35 E5 cases, and 19/20 SPLADE cases. These conditional counts are descriptive and are based on small subgroups. Relative to all 6,484 HotpotQA queries supported by BM25 at $k{=}10$, these flip-associated answer regressions correspond to 0.37\% for BGE, 0.51\% for E5, and 0.29\% for SPLADE. The check uses one reader, one dataset, and $k{=}10$. It records an association in this setting and is not a general estimate of downstream harm.

\begin{table}[t]
\caption{HotpotQA answer check at $k{=}10$. ``Regression'' denotes a query answered correctly from BM25 passages and incorrectly from replacement passages.}
\label{tab:answers}
\centering
\small
\begin{tabular}{llrrrrr}
\toprule
Model & Group & Read & Regression & Preserved & Improved & Both wrong \\
\midrule
BGE & Flip & 143 & 24 & 0 & 3 & 116 \\
BGE & Control & 143 & 9 & 20 & 14 & 100 \\
E5 & Flip & 156 & 33 & 2 & 3 & 118 \\
E5 & Control & 156 & 2 & 42 & 14 & 98 \\
SPLADE & Flip & 125 & 19 & 1 & 4 & 101 \\
SPLADE & Control & 125 & 9 & 23 & 6 & 87 \\
\bottomrule
\end{tabular}
\end{table}

\begin{figure}[t]
\centering
\includegraphics[width=\linewidth]{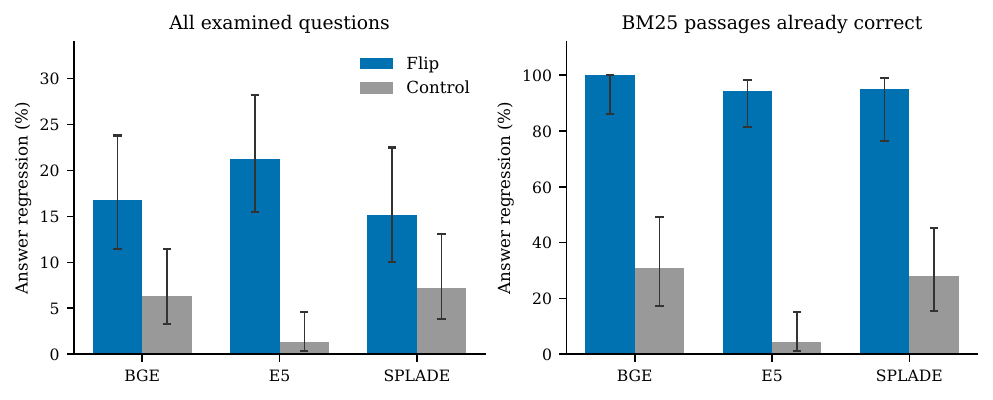}
\caption{HotpotQA answer regressions at $k{=}10$. The left panel includes all retrieval flips and matched controls. The right panel conditions on queries for which the BM25 passages already produced the correct answer. Bars show 95\% Wilson intervals.}
\label{fig:answers}
\end{figure}

\section{Discussion}

Every replacement has positive net gain in all 45 model--corpus depth conditions, and every condition also contains negative flips. Aggregate retrieval quality records whether the replacement supports more queries overall. Compatibility records whether previously successful queries remain supported. The two summaries describe different aspects of a retriever update. Negative-flip rates are lower in the larger-depth settings, where the BM25-supported cohort is defined separately at each $k$, and the rates are nonzero at $k{=}50$. One-parameter changes inside BM25 break fewer reference successes than the replacements at every evaluated depth. On HotpotQA, requiring every positive qrel passage in the top $k$ produces higher negative-flip rates on a smaller BM25-supported cohort, with positive net gain at every depth whose denominator is nonzero. The union baseline has zero retrieval negative flips by construction, because it retains every BM25 top-$k$ passage, at a context size of up to $2k$. Fixed-budget combinations reduce broken queries, with net-gain trade-offs that vary by corpus, retriever, and combination method. Compatibility also varies by corpus. E5 preserves more BM25 successes than BGE on Natural Questions. The two dense retrievers are not detectably different on HotpotQA, and the ordering reverses at larger depths on FiQA. The evaluated SPLADE configuration shows the same general depth pattern. These findings agree with prior evidence that model updates can improve aggregate performance and change outcomes for individual examples \citep{milanifard2016launch,bahri2021locally}. The HotpotQA analysis shows that retrieval-level and answer-level regressions are related and distinct. Because the analysis uses one reader and one dataset, it should be interpreted as a limited downstream check. RAGFlip can serve as an additional pre-deployment diagnostic by identifying fixed, preserved, and broken queries. It is not itself a deployment decision rule. The accompanying RAGFlip implementation applies the same before/after analysis to user-provided retrieval traces and identifies fixed, preserved, and broken queries.

\section{Limitations}

The primary support label requires at least one positively judged passage in the top $k$, while the strict HotpotQA check requires all positive qrel passages. Because BEIR judgments may be incomplete, a qrel-defined negative flip does not necessarily imply a genuine loss of relevant evidence. The reader analysis is also limited to one frozen 7B model, HotpotQA, and $k{=}10$, with relatively few BM25-correct flip queries. Its results should therefore be interpreted as a limited downstream check. FiQA contains only 648 test queries, which leads to wider confidence intervals. SPLADE inputs are truncated to 256 tokens, and the main configuration retains the 256 largest positive weights. Removing this weight cutoff produced nearly identical transition counts. BM25 results depend on the fixed tokenization, stopword list, and parameter settings used in this study. Comparisons across retrieval depths use different BM25-supported cohorts at each $k$. The union baseline has zero retrieval flips by construction, but this does not imply zero answer regressions. And retriever training data differ. E5 includes Natural Questions during supervised fine-tuning \citep{wang2022e5}, so its Natural Questions results should not be interpreted as purely zero-shot.

\section{Conclusion}

Aggregate retrieval improvements can coexist with query-level regressions relative to the existing retriever. We present \textsc{RAGFlip}, a query-level framework for evaluating behavioral compatibility during retriever replacement. Across three BEIR corpora, two dense retrievers, the evaluated SPLADE configuration, and five retrieval depths, every replacement produces positive net gain and also introduces negative flips. Negative-flip rates are lower in the larger-depth settings, where the BM25-supported cohort is defined separately at each $k$. Their magnitude and ordering depend on the corpus and retriever. Fixed-budget combinations reduce broken queries, with net-gain trade-offs that vary by setting. Retaining both rankings eliminates qrel-defined retrieval flips by construction, because every BM25 top-$k$ passage is kept, at the cost of a larger context. The HotpotQA reader analysis provides a limited downstream check and shows that some retrieval regressions coincide with answer regressions. Reporting query-level compatibility alongside aggregate retrieval quality gives a more complete view of a retriever replacement. Future work can extend the analysis to additional retrievers, datasets, and reader models.

\subsubsection*{Broader Impact Statement}

This study uses public question-answering benchmarks and does not evaluate people or target high-stakes deployment. Its main implication is that aggregate gains can hide regressions on previously successful queries. Query-level compatibility checks can identify these regressions. They are not a deployment decision rule. The evidence is limited to benchmark queries with existing relevance judgments.

\bibliography{main}

@inproceedings{karpukhin2020dense,
  title = {Dense Passage Retrieval for Open-Domain Question Answering},
  author = {Karpukhin, Vladimir and Oguz, Barlas and Min, Sewon and Lewis, Patrick and Wu, Ledell and Edunov, Sergey and Chen, Danqi and Yih, Wen-tau},
  booktitle = {Proceedings of the 2020 Conference on Empirical Methods in Natural Language Processing (EMNLP)},
  pages = {6769--6781},
  year = {2020},
  publisher = {Association for Computational Linguistics},
  doi = {10.18653/v1/2020.emnlp-main.550},
  url = {https://aclanthology.org/2020.emnlp-main.550/}
}

@inproceedings{thakur2021beir,
  title = {{BEIR}: A Heterogeneous Benchmark for Zero-shot Evaluation of Information Retrieval Models},
  author = {Thakur, Nandan and Reimers, Nils and R{\"u}ckl{\'e}, Andreas and Srivastava, Abhishek and Gurevych, Iryna},
  booktitle = {Thirty-Fifth Conference on Neural Information Processing Systems Datasets and Benchmarks Track},
  year = {2021},
  url = {https://openreview.net/forum?id=wCu6T5xFjeJ}
}

@article{wang2022e5,
  title = {Text Embeddings by Weakly-Supervised Contrastive Pre-training},
  author = {Wang, Liang and Yang, Nan and Huang, Xiaolong and Jiao, Binxing and Yang, Linjun and Jiang, Daxin and Majumder, Rangan and Wei, Furu},
  journal = {arXiv preprint arXiv:2212.03533},
  year = {2022},
  url = {https://arxiv.org/abs/2212.03533}
}

@inproceedings{yan2021positive,
  title = {Positive-Congruent Training: Towards Regression-Free Model Updates},
  author = {Yan, Sijie and Xiong, Yuanjun and Kundu, Kaustav and Yang, Shuo and Deng, Siqi and Wang, Meng and Xia, Wei and Soatto, Stefano},
  booktitle = {Proceedings of the IEEE/CVF Conference on Computer Vision and Pattern Recognition},
  pages = {14299--14308},
  year = {2021},
  doi = {10.1109/CVPR46437.2021.01407}
}

@inproceedings{cai2023l2r,
  title = {{L2R}: Lifelong Learning for First-stage Retrieval with Backward-Compatible Representations},
  author = {Cai, Yinqiong and Bi, Keping and Fan, Yixing and Guo, Jiafeng and Chen, Wei and Cheng, Xueqi},
  booktitle = {Proceedings of the 32nd ACM International Conference on Information and Knowledge Management},
  pages = {183--192},
  year = {2023},
  publisher = {ACM},
  doi = {10.1145/3583780.3614947},
  url = {https://doi.org/10.1145/3583780.3614947}
}

@article{xiao2023cpack,
  title = {{C-Pack}: Packed Resources For General Chinese Embeddings},
  author = {Xiao, Shitao and Liu, Zheng and Zhang, Peitian and Muennighoff, Niklas and Lian, Defu and Nie, Jian-Yun},
  journal = {arXiv preprint arXiv:2309.07597},
  year = {2023},
  url = {https://arxiv.org/abs/2309.07597}
}

@inproceedings{formal2021splade,
  title = {{SPLADE}: Sparse Lexical and Expansion Model for First Stage Ranking},
  author = {Formal, Thibault and Piwowarski, Benjamin and Clinchant, St{\'e}phane},
  booktitle = {Proceedings of the 44th International ACM SIGIR Conference on Research and Development in Information Retrieval},
  pages = {2288--2292},
  year = {2021},
  publisher = {ACM},
  doi = {10.1145/3404835.3463098},
  url = {https://doi.org/10.1145/3404835.3463098}
}

@inproceedings{shen2020backward,
  title = {Towards Backward-Compatible Representation Learning},
  author = {Shen, Yantao and Xiong, Yuanjun and Xia, Wei and Soatto, Stefano},
  booktitle = {Proceedings of the IEEE/CVF Conference on Computer Vision and Pattern Recognition},
  pages = {6368--6377},
  year = {2020}
}

@inproceedings{schumann2024backward,
  title = {Backward Compatibility During Data Updates by Weight Interpolation},
  author = {Schumann, Raphael and Mansimov, Elman and Lai, Yi-An and Pappas, Nikolaos and Gao, Xibin and Zhang, Yi},
  booktitle = {Proceedings of the 18th Conference of the European Chapter of the Association for Computational Linguistics},
  pages = {2846--2861},
  year = {2024},
  publisher = {Association for Computational Linguistics},
  doi = {10.18653/v1/2024.eacl-long.174}
}

@inproceedings{xiong2021ance,
  title = {Approximate Nearest Neighbor Negative Contrastive Learning for Dense Text Retrieval},
  author = {Xiong, Lee and Xiong, Chenyan and Li, Ye and Tang, Kwok-Fung and Liu, Jialin and Bennett, Paul N. and Ahmed, Junaid and Overwijk, Arnold},
  booktitle = {International Conference on Learning Representations},
  year = {2021}
}

@inproceedings{khattab2020colbert,
  title = {{ColBERT}: Efficient and Effective Passage Search via Contextualized Late Interaction over {BERT}},
  author = {Khattab, Omar and Zaharia, Matei},
  booktitle = {Proceedings of the 43rd International ACM SIGIR Conference on Research and Development in Information Retrieval},
  pages = {39--48},
  year = {2020},
  publisher = {ACM},
  doi = {10.1145/3397271.3401075}
}

@inproceedings{gao2021coil,
  title = {{COIL}: Revisit Exact Lexical Match in Information Retrieval with Contextualized Inverted List},
  author = {Gao, Luyu and Dai, Zhuyun and Callan, Jamie},
  booktitle = {Proceedings of the 2021 Conference of the North American Chapter of the Association for Computational Linguistics: Human Language Technologies},
  pages = {3030--3042},
  year = {2021},
  publisher = {Association for Computational Linguistics},
  doi = {10.18653/v1/2021.naacl-main.241}
}

@inproceedings{wang2021interpolation,
  title = {{BERT}-based Dense Retrievers Require Interpolation with {BM25} for Effective Passage Retrieval},
  author = {Wang, Shuai and Zhuang, Shengyao and Zuccon, Guido},
  booktitle = {Proceedings of the 2021 ACM SIGIR International Conference on Theory of Information Retrieval},
  pages = {317--324},
  year = {2021},
  publisher = {ACM},
  doi = {10.1145/3471158.3472233}
}

@inproceedings{cormack2009rrf,
  title = {Reciprocal Rank Fusion Outperforms Condorcet and Individual Rank Learning Methods},
  author = {Cormack, Gordon V. and Clarke, Charles L. A. and B{\"u}ttcher, Stefan},
  booktitle = {Proceedings of the 32nd International ACM SIGIR Conference on Research and Development in Information Retrieval},
  pages = {758--759},
  year = {2009},
  publisher = {ACM},
  doi = {10.1145/1571941.1572114}
}

@inproceedings{craswell2020trec,
  title = {Overview of the {TREC} 2020 Deep Learning Track},
  author = {Craswell, Nick and Mitra, Bhaskar and Yilmaz, Emine and Campos, Daniel},
  booktitle = {Proceedings of the Twenty-Ninth Text REtrieval Conference (TREC 2020)},
  year = {2020},
  publisher = {National Institute of Standards and Technology},
  doi = {10.6028/NIST.SP.1266.deep-overview}
}

@inproceedings{wang2012robust,
  title = {Robust Ranking Models via Risk-Sensitive Optimization},
  author = {Wang, Lidan and Bennett, Paul N. and Collins-Thompson, Kevyn},
  booktitle = {Proceedings of the 35th International ACM SIGIR Conference on Research and Development in Information Retrieval},
  pages = {761--770},
  year = {2012},
  publisher = {ACM},
  doi = {10.1145/2348283.2348385},
  url = {https://doi.org/10.1145/2348283.2348385}
}

@inproceedings{kwiatkowski2019natural,
  title = {Natural Questions: A Benchmark for Question Answering Research},
  author = {Kwiatkowski, Tom and Palomaki, Jennimaria and Redfield, Olivia and Collins, Michael and Parikh, Ankur and Alberti, Chris and Epstein, Danielle and Polosukhin, Illia and Devlin, Jacob and Lee, Kenton and Toutanova, Kristina and Jones, Llion and Kelcey, Ming-Wei and Chang, Ming-Wei and Dai, Andrew M. and Uszkoreit, Jakob and Le, Quoc and Petrov, Slav},
  booktitle = {Transactions of the Association for Computational Linguistics},
  volume = {7},
  pages = {453--466},
  year = {2019},
  doi = {10.1162/tacl_a_00276},
  url = {https://aclanthology.org/Q19-1026/}
}

@inproceedings{yang2018hotpotqa,
  title = {HotpotQA: A Dataset for Diverse, Explainable Multi-hop Question Answering},
  author = {Yang, Zhilin and Qi, Peng and Zhang, Saizheng and Bengio, Yoshua and Cohen, William W. and Salakhutdinov, Ruslan and Manning, Christopher D.},
  booktitle = {Proceedings of the 2018 Conference on Empirical Methods in Natural Language Processing},
  pages = {2369--2380},
  year = {2018},
  publisher = {Association for Computational Linguistics},
  doi = {10.18653/v1/D18-1259},
  url = {https://aclanthology.org/D18-1259/}
}

@inproceedings{maia2018fiqa,
  title = {{WWW}'18 Open Challenge: Financial Opinion Mining and Question Answering},
  author = {Maia, Macedo and Handschuh, Siegfried and Freitas, Andr{\'e} and Davis, Brian and McDermott, Ross and Zarrouk, Manel and Balahur, Alexandra},
  booktitle = {Companion Proceedings of The Web Conference 2018},
  pages = {1941--1942},
  year = {2018},
  publisher = {ACM},
  doi = {10.1145/3184558.3192301},
  url = {https://doi.org/10.1145/3184558.3192301}
}

@article{robertson2009probabilistic,
  title = {The Probabilistic Relevance Framework: {BM25} and Beyond},
  author = {Robertson, Stephen and Zaragoza, Hugo},
  journal = {Foundations and Trends in Information Retrieval},
  volume = {3},
  number = {4},
  pages = {333--389},
  year = {2009},
  publisher = {Now Publishers},
  doi = {10.1561/1500000019}
}

@inproceedings{formal2022splade,
  title = {From Distillation to Hard Negative Sampling: Making Sparse Neural {IR} Models More Effective},
  author = {Formal, Thibault and Lassance, Carlos and Piwowarski, Benjamin and Clinchant, St{\'e}phane},
  booktitle = {Proceedings of the 45th International ACM SIGIR Conference on Research and Development in Information Retrieval},
  pages = {2353--2359},
  year = {2022},
  publisher = {ACM},
  doi = {10.1145/3477495.3531857},
  url = {https://doi.org/10.1145/3477495.3531857}
}

@article{wilson1927probable,
  title = {Probable Inference, the Law of Succession, and Statistical Inference},
  author = {Wilson, Edwin B.},
  journal = {Journal of the American Statistical Association},
  volume = {22},
  number = {158},
  pages = {209--212},
  year = {1927},
  doi = {10.1080/01621459.1927.10502953},
  url = {https://doi.org/10.1080/01621459.1927.10502953}
}

@article{mcnemar1947note,
  title = {Note on the Sampling Error of the Difference Between Correlated Proportions or Percentages},
  author = {McNemar, Quinn},
  journal = {Psychometrika},
  volume = {12},
  number = {2},
  pages = {153--157},
  year = {1947},
  doi = {10.1007/BF02295996},
  url = {https://doi.org/10.1007/BF02295996}
}

@article{holm1979simple,
  title = {A Simple Sequentially Rejective Multiple Test Procedure},
  author = {Holm, Sture},
  journal = {Scandinavian Journal of Statistics},
  volume = {6},
  number = {2},
  pages = {65--70},
  year = {1979},
  doi = {10.2307/4615733},
  url = {https://doi.org/10.2307/4615733}
}

@article{qwen2025qwen25,
  title = {Qwen2.5 Technical Report},
  author = {Yang, An and Yang, Baosong and Zhang, Beichen and Hui, Binyuan and Zheng, Bo and Yu, Bowen and Li, Chengyuan and Liu, Dayiheng and Huang, Fei and Wei, Haoran and Lin, Huan and Yang, Jian and Tu, Jianhong and Zhang, Jianwei and Yang, Jianxin and Yang, Jiaxi and Zhou, Jingren and Lin, Junyang and Dang, Kai and Lu, Keming and Bao, Keqin and Yang, Kexin and Yu, Le and Li, Mei and Xue, Mingfeng and Zhang, Pei and Zhu, Qin and Men, Rui and Lin, Runji and Li, Tianhao and Tang, Tianyi and Xia, Tingyu and Ren, Xingzhang and Ren, Xuancheng and Fan, Yang and Su, Yang and Zhang, Yichang and Wan, Yu and Liu, Yuqiong and Cui, Zeyu and Zhang, Zhenru and Qiu, Zihan},
  journal = {arXiv preprint arXiv:2412.15115},
  year = {2024},
  url = {https://arxiv.org/abs/2412.15115}
}

@inproceedings{milanifard2016launch,
  title = {Launch and Iterate: Reducing Prediction Churn},
  author = {Milani Fard, Mahdi and Cormier, Quentin and Canini, Kevin and Gupta, Maya},
  booktitle = {Advances in Neural Information Processing Systems},
  volume = {29},
  pages = {3179--3187},
  year = {2016}
}

@inproceedings{bahri2021locally,
  title = {Locally Adaptive Label Smoothing Improves Predictive Churn},
  author = {Bahri, Dara and Jiang, Heinrich},
  booktitle = {Proceedings of the 38th International Conference on Machine Learning},
  volume = {139},
  series = {Proceedings of Machine Learning Research},
  pages = {532--542},
  year = {2021},
  publisher = {PMLR},
  url = {https://proceedings.mlr.press/v139/bahri21a.html}
}

@inproceedings{lai2023improving,
  title = {Improving Prediction Backward-Compatiblility in NLP Model Upgrade with Gated Fusion},
  author = {Lai, Yi-An and Mansimov, Elman and Xie, Yuqing and Zhang, Yi},
  booktitle = {Findings of the Association for Computational Linguistics: EACL 2023},
  pages = {1010--1022},
  year = {2023},
  publisher = {Association for Computational Linguistics},
  doi = {10.18653/v1/2023.findings-eacl.74},
  url = {https://aclanthology.org/2023.findings-eacl.74/}
}

@misc{irankhah2026archehrqa,
  title        = {Yale-DM-Lab at ArchEHR-QA 2026: Deterministic Grounding and Multi-Pass Evidence Alignment for EHR Question Answering},
  author       = {Irankhah, Elyas and Fodeh, Samah},
  year         = {2026},
  eprint       = {2604.07116},
  archivePrefix = {arXiv},
  primaryClass = {cs.CL},
  url          = {https://arxiv.org/abs/2604.07116}
}

@inproceedings{srivastava2020backward,
  title     = {An Empirical Analysis of Backward Compatibility in Machine Learning Systems},
  author    = {Srivastava, Megha and Nushi, Besmira and Kamar, Ece and Shah, Shital and Horvitz, Eric},
  booktitle = {Proceedings of the 26th ACM SIGKDD International Conference on Knowledge Discovery \& Data Mining},
  year      = {2020}
}
\bibliographystyle{tmlr}

\appendix

\section{Supplementary Materials}
\label{app:supplementary}

This appendix provides the complete statistical and depth-wise results that support the analyses reported in the main text. It includes the full paired significance tests, the raw query-level transition counts, the fixed-budget combination results across all evaluated retrieval depths, a check that the saved BM25 rankings and primary HotpotQA counts match the stated configuration, the within-BM25 parameter comparison, the strict HotpotQA support check, and a SPLADE weight-cutoff recomputation. Figure~\ref{fig:churn} plots the $k{=}10$ broken counts, and Figure~\ref{fig:strict} plots the HotpotQA rates under both support labels.

\subsection{Paired Tests Across Retrieval Depths}
\label{app:paired}

Table~\ref{tab:paired-all} extends the paired BGE--E5 comparison in Table~\ref{tab:paired} to all five retrieval depths. The analysis is restricted to queries supported by BM25 at each depth. ``Only BGE'' denotes queries that become negative flips under BGE and remain supported under E5, while ``Only E5'' denotes the reverse case. The exact McNemar test evaluates whether these discordant counts differ, with Holm correction applied across the five depth-wise tests within each corpus.

\begin{table}[h]
\caption{Exact McNemar tests comparing BGE and E5 negative flips on BM25-supported queries across retrieval depths. Holm-adjusted $p$-values correct for the five depth-wise comparisons within each corpus.}
\label{tab:paired-all}
\centering
\begin{tabular}{llrrrr}
\toprule
Corpus & $k$ & Only BGE & Only E5 & $p$ & $p_{\mathrm{Holm}}$ \\
\midrule
Natural Questions & 1  & 84 & 38 & $3.8 \times 10^{-5}$ & 0.0002 \\
Natural Questions & 5  & 82 & 38 & $7.3 \times 10^{-5}$ & 0.0002 \\
Natural Questions & 10 & 60 & 26 & 0.0003 & 0.0006 \\
Natural Questions & 20 & 59 & 17 & $1.4 \times 10^{-6}$ & $7.0 \times 10^{-6}$ \\
Natural Questions & 50 & 33 & 15 & 0.013 & 0.013 \\
\midrule
HotpotQA & 1  & 226 & 220 & 0.81 & 1.00 \\
HotpotQA & 5  & 106 & 88  & 0.22 & 1.00 \\
HotpotQA & 10 & 64  & 77  & 0.31 & 1.00 \\
HotpotQA & 20 & 50  & 51  & 1.00 & 1.00 \\
HotpotQA & 50 & 33  & 38  & 0.64 & 1.00 \\
\midrule
FiQA & 1  & 10 & 15 & 0.42 & 0.42 \\
FiQA & 5  & 10 & 23 & 0.035 & 0.11 \\
FiQA & 10 & 10 & 20 & 0.099 & 0.197 \\
FiQA & 20 & 5  & 24 & 0.0005 & 0.0027 \\
FiQA & 50 & 3  & 17 & 0.0026 & 0.010 \\
\bottomrule
\end{tabular}
\end{table}

\subsection{Complete Transition Counts}
\label{app:counts}

Tables~\ref{tab:counts-nq}--\ref{tab:counts-fiqa} report the complete query-level transition counts underlying the rates and net gains presented in the main text. ``Both'' denotes queries supported by BM25 and the replacement retriever, ``Fixed'' denotes BM25 failures recovered by the replacement, and ``Broken'' denotes BM25 successes that become negative flips. For each row, New $=$ Both $+$ Fixed, BM25 $=$ Both $+$ Broken, and Net $=$ Fixed $-$ Broken.

\begin{table}[h]
\caption{Natural Questions transition counts across retrieval depths.}
\label{tab:counts-nq}
\centering
\small
\begin{tabular}{lrrrrrr}
\toprule
Model & $k$ & BM25 & New & Both & Fixed & Broken \\
\midrule
BGE & 1  & 517  & 1225 & 323  & 902  & 194 \\
BGE & 5  & 1240 & 2397 & 1091 & 1306 & 149 \\
BGE & 10 & 1600 & 2777 & 1503 & 1274 & 97 \\
BGE & 20 & 1956 & 3012 & 1867 & 1145 & 89 \\
BGE & 50 & 2347 & 3211 & 2292 & 919  & 55 \\
E5  & 1  & 517  & 1476 & 369  & 1107 & 148 \\
E5  & 5  & 1240 & 2646 & 1135 & 1511 & 105 \\
E5  & 10 & 1600 & 2959 & 1537 & 1422 & 63 \\
E5  & 20 & 1956 & 3140 & 1909 & 1231 & 47 \\
E5  & 50 & 2347 & 3263 & 2310 & 953  & 37 \\
SPLADE & 1  & 517  & 1195 & 349  & 846  & 168 \\
SPLADE & 5  & 1240 & 2329 & 1119 & 1210 & 121 \\
SPLADE & 10 & 1600 & 2648 & 1511 & 1137 & 89 \\
SPLADE & 20 & 1956 & 2933 & 1897 & 1036 & 59 \\
SPLADE & 50 & 2347 & 3154 & 2309 & 845  & 38 \\
\bottomrule
\end{tabular}
\end{table}

\begin{table}[h]
\caption{HotpotQA transition counts across retrieval depths.}
\label{tab:counts-hotpot}
\centering
\small
\begin{tabular}{lrrrrrr}
\toprule
Model & $k$ & BM25 & New & Both & Fixed & Broken \\
\midrule
BGE & 1  & 4875 & 6050 & 4377 & 1673 & 498 \\
BGE & 5  & 6130 & 6940 & 5925 & 1015 & 205 \\
BGE & 10 & 6484 & 7088 & 6341 & 747  & 143 \\
BGE & 20 & 6759 & 7187 & 6654 & 533  & 105 \\
BGE & 50 & 7025 & 7282 & 6964 & 318  & 61 \\
E5  & 1  & 4875 & 6075 & 4383 & 1692 & 492 \\
E5  & 5  & 6130 & 6959 & 5943 & 1016 & 187 \\
E5  & 10 & 6484 & 7073 & 6328 & 745  & 156 \\
E5  & 20 & 6759 & 7177 & 6653 & 524  & 106 \\
E5  & 50 & 7025 & 7270 & 6959 & 311  & 66 \\
SPLADE & 1  & 4875 & 6046 & 4458 & 1588 & 417 \\
SPLADE & 5  & 6130 & 6915 & 5969 & 946  & 161 \\
SPLADE & 10 & 6484 & 7067 & 6359 & 708  & 125 \\
SPLADE & 20 & 6759 & 7183 & 6674 & 509  & 85 \\
SPLADE & 50 & 7025 & 7279 & 6973 & 306 & 52 \\
\bottomrule
\end{tabular}
\end{table}

\begin{table}[h]
\caption{FiQA transition counts across retrieval depths.}
\label{tab:counts-fiqa}
\centering
\small
\begin{tabular}{lrrrrrr}
\toprule
Model & $k$ & BM25 & New & Both & Fixed & Broken \\
\midrule
BGE & 1  & 139 & 288 & 112 & 176 & 27 \\
BGE & 5  & 246 & 418 & 221 & 197 & 25 \\
BGE & 10 & 306 & 462 & 278 & 184 & 28 \\
BGE & 20 & 365 & 515 & 349 & 166 & 16 \\
BGE & 50 & 414 & 556 & 404 & 152 & 10 \\
E5  & 1  & 139 & 269 & 107 & 162 & 32 \\
E5  & 5  & 246 & 378 & 208 & 170 & 38 \\
E5  & 10 & 306 & 433 & 268 & 165 & 38 \\
E5  & 20 & 365 & 477 & 330 & 147 & 35 \\
E5  & 50 & 414 & 530 & 390 & 140 & 24 \\
SPLADE & 1  & 139 & 216 & 114 & 102 & 25 \\
SPLADE & 5  & 246 & 349 & 219 & 130 & 27 \\
SPLADE & 10 & 306 & 395 & 278 & 117 & 28 \\
SPLADE & 20 & 365 & 442 & 332 & 110 & 33 \\
SPLADE & 50 & 414 & 502 & 395 & 107 & 19 \\
\bottomrule
\end{tabular}
\end{table}

\subsection{Fixed-Budget Combination Results}
\label{app:fusion}

Tables~\ref{tab:fusion-nq}--\ref{tab:fusion-fiqa} extend the fixed-budget analysis in Table~\ref{tab:mix} to every retrieval depth. The alternate and RRF methods each return exactly $k$ distinct passages, so their regression behavior can be compared with the replacement retriever under the same retrieval budget. At $k{=}1$, alternation begins with the replacement retriever and is therefore identical to the replacement ranking. The tables report the number of broken BM25 successes and the resulting net gain for each combination.

\begin{table}[h]
\caption{Natural Questions fixed-budget combination results across retrieval depths.}
\label{tab:fusion-nq}
\centering
\small
\begin{tabular}{lrrrrr}
\toprule
Model & $k$ & Alternate broken & Alternate net & RRF broken & RRF net \\
\midrule
BGE & 1  & 194 & +708  & 66 & +423 \\
BGE & 5  & 83  & +996  & 34 & +807 \\
BGE & 10 & 48  & +1024 & 23 & +873 \\
BGE & 20 & 45  & +960  & 20 & +942 \\
BGE & 50 & 25  & +830  & 17 & +834 \\
E5  & 1  & 148 & +959  & 46 & +521 \\
E5  & 5  & 53  & +1224 & 23 & +878 \\
E5  & 10 & 32  & +1205 & 12 & +997 \\
E5  & 20 & 24  & +1103 & 11 & +1050 \\
E5  & 50 & 14  & +892  & 9  & +894 \\
SPLADE & 1  & 168 & +678 & 47 & +408 \\
SPLADE & 5  & 81  & +933 & 42 & +656 \\
SPLADE & 10 & 40  & +937 & 13 & +726 \\
SPLADE & 20 & 34  & +858 & 14 & +761 \\
SPLADE & 50 & 13  & +757 & 8  & +755 \\
\bottomrule
\end{tabular}
\end{table}

\begin{table}[h]
\caption{HotpotQA fixed-budget combination results across retrieval depths.}
\label{tab:fusion-hotpot}
\centering
\small
\begin{tabular}{lrrrrr}
\toprule
Model & $k$ & Alternate broken & Alternate net & RRF broken & RRF net \\
\midrule
BGE & 1  & 498 & +1175 & 185 & +1032 \\
BGE & 5  & 50  & +896  & 37  & +832 \\
BGE & 10 & 40  & +669  & 24  & +677 \\
BGE & 20 & 20  & +482  & 13  & +500 \\
BGE & 50 & 8   & +288  & 7   & +288 \\
E5  & 1  & 492 & +1200 & 198 & +938 \\
E5  & 5  & 44  & +912  & 56  & +780 \\
E5  & 10 & 37  & +671  & 31  & +645 \\
E5  & 20 & 21  & +477  & 15  & +489 \\
E5  & 50 & 13  & +279  & 9   & +282 \\
SPLADE & 1  & 417 & +1171 & 158 & +890 \\
SPLADE & 5  & 56  & +821  & 58  & +655 \\
SPLADE & 10 & 45  & +615  & 41  & +542 \\
SPLADE & 20 & 23  & +447  & 17  & +446 \\
SPLADE & 50 & 12  & +278  & 10  & +279 \\
\bottomrule
\end{tabular}
\end{table}

\begin{table}[h]
\caption{FiQA fixed-budget combination results across retrieval depths.}
\label{tab:fusion-fiqa}
\centering
\small
\begin{tabular}{lrrrrr}
\toprule
Model & $k$ & Alternate broken & Alternate net & RRF broken & RRF net \\
\midrule
BGE & 1  & 27 & +149 & 15 & +81 \\
BGE & 5  & 11 & +157 & 5  & +140 \\
BGE & 10 & 17 & +143 & 6  & +137 \\
BGE & 20 & 7  & +133 & 4  & +127 \\
BGE & 50 & 4  & +131 & 4  & +128 \\
E5  & 1  & 32 & +130 & 13 & +86 \\
E5  & 5  & 16 & +130 & 14 & +131 \\
E5  & 10 & 20 & +115 & 15 & +120 \\
E5  & 20 & 15 & +113 & 10 & +117 \\
E5  & 50 & 7  & +110 & 3  & +114 \\
SPLADE & 1  & 25 & +77 & 17 & +45 \\
SPLADE & 5  & 14 & +83 & 12 & +70 \\
SPLADE & 10 & 15 & +84 & 8  & +69 \\
SPLADE & 20 & 22 & +66 & 13 & +68 \\
SPLADE & 50 & 5  & +89 & 4  & +89 \\
\bottomrule
\end{tabular}
\end{table}

\subsection{Reference Ranking Check}
\label{app:reproduction}

The saved BM25 top-50 rankings were recomputed from the BEIR corpora with the tokenizer, the 43-word stop list, and the parameters in Section~\ref{sec:setup} ($k_1{=}1.5$, $b{=}0.75$). Document order follows the corpus file order used for the saved rankings. The recomputed top-50 list was identical to the saved ranking for every query on Natural Questions, HotpotQA, and FiQA.

The primary HotpotQA counts were recomputed from the saved rankings under the same any-relevant definition used in the main text: a query is supported when at least one positive qrel passage appears in the top $k$. The recomputed broken, fixed, and both counts match Table~\ref{tab:counts-hotpot} at all five depths.

\subsection{Within-BM25 Parameter Comparison}
\label{app:churn}

\begin{figure}[h]
\centering
\includegraphics[width=\linewidth]{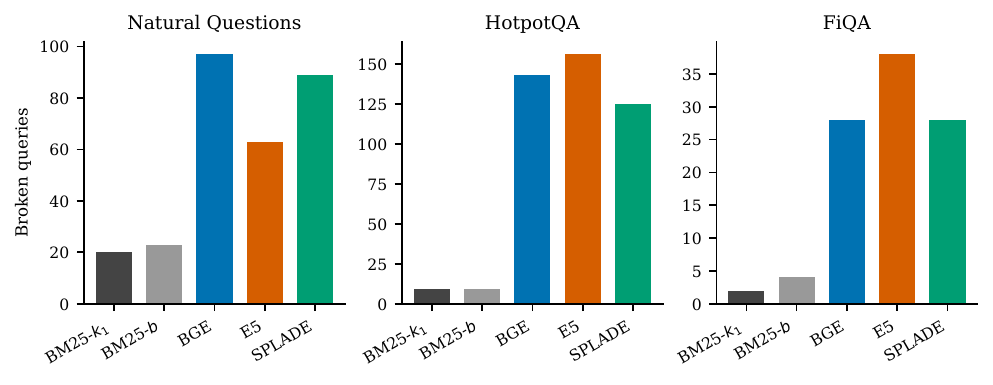}
\caption{Broken queries at $k{=}10$. BM25-$k_1$ uses $k_1{=}1.2$ and $b{=}0.75$. BM25-$b$ uses $k_1{=}1.5$ and $b{=}0.65$. Variant counts are from Table~\ref{tab:churn}. Replacement counts are the broken cells of Table~\ref{tab:transitions}. Each panel has its own vertical scale.}
\label{fig:churn}
\end{figure}

Table~\ref{tab:churn} compares the reference BM25 ranking with two one-parameter variants. Tokenization, the stop list, the corpora, and the qrels stay the same. BM25-$k_1$ changes $k_1$ from 1.5 to 1.2. BM25-$b$ changes $b$ from 0.75 to 0.65. The recomputed reference top-50 list matched the saved ranking on every query, so these counts use the same reference as the main tables. A broken query is supported by the reference and missed by the variant. A fixed query is missed by the reference and supported by the variant. Figure~\ref{fig:churn} places the $k{=}10$ broken counts next to the three replacements.

\begin{table}[h]
\caption{Within-BM25 transitions relative to $k_1{=}1.5$, $b{=}0.75$. BM25-$k_1$ uses $k_1{=}1.2$. BM25-$b$ uses $b{=}0.65$.}
\label{tab:churn}
\centering
\small
\begin{tabular}{llrrrrr}
\toprule
Corpus & $k$ & BM25 & $k_1{=}1.2$ broken & $k_1{=}1.2$ fixed & $b{=}0.65$ broken & $b{=}0.65$ fixed \\
\midrule
Natural Questions & 1  & 517  & 19 & 31  & 30 & 63 \\
Natural Questions & 5  & 1240 & 22 & 45  & 23 & 85 \\
Natural Questions & 10 & 1600 & 20 & 55  & 23 & 100 \\
Natural Questions & 20 & 1956 & 18 & 51  & 15 & 88 \\
Natural Questions & 50 & 2347 & 21 & 44  & 11 & 60 \\
\midrule
HotpotQA & 1  & 4875 & 28 & 176 & 31 & 212 \\
HotpotQA & 5  & 6130 & 17 & 141 & 20 & 165 \\
HotpotQA & 10 & 6484 & 9  & 108 & 9  & 134 \\
HotpotQA & 20 & 6759 & 10 & 82  & 6  & 89 \\
HotpotQA & 50 & 7025 & 8  & 39  & 3  & 42 \\
\midrule
FiQA & 1  & 139 & 7 & 7  & 7 & 9 \\
FiQA & 5  & 246 & 6 & 6  & 6 & 13 \\
FiQA & 10 & 306 & 2 & 10 & 4 & 12 \\
FiQA & 20 & 365 & 4 & 7  & 7 & 5 \\
FiQA & 50 & 414 & 4 & 7  & 4 & 8 \\
\bottomrule
\end{tabular}
\end{table}

\subsection{Strict HotpotQA Support}
\label{app:strict}

\begin{figure}[h]
\centering
\includegraphics[width=\linewidth]{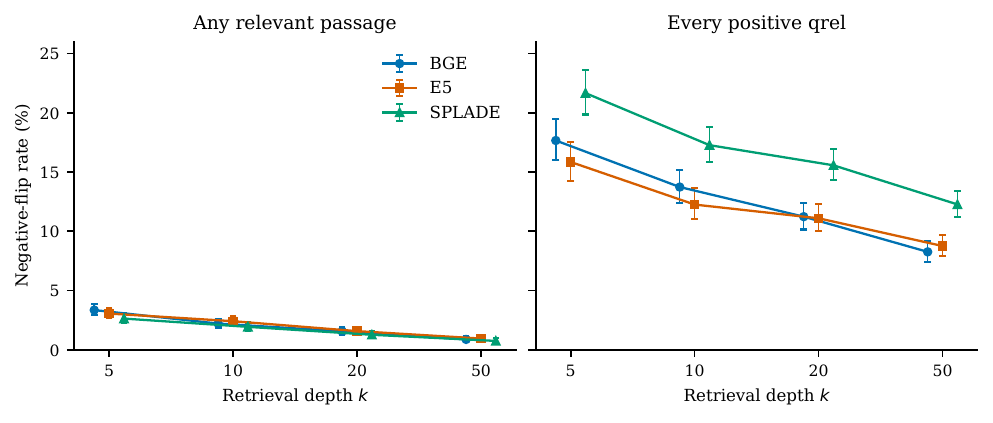}
\caption{HotpotQA negative-flip rates. The left panel requires one positive qrel passage. The right panel requires every positive qrel passage. Both panels use the same vertical scale. $k{=}1$ is omitted because the strict BM25-supported count is zero. Bars show 95\% Wilson intervals. Counts are in Tables~\ref{tab:counts-hotpot} and \ref{tab:strict}.}
\label{fig:strict}
\end{figure}

Table~\ref{tab:strict} repeats the HotpotQA replacement comparison when a query is supported only if every positive qrel passage appears in the top $k$. The rankings are the saved rankings used in the main tables. The any-relevant counts from those rankings match Table~\ref{tab:counts-hotpot} at all five depths. At $k{=}1$, BM25 supports no query under the strict label, so the negative-flip rate is undefined. At the other depths the BM25-supported cohort is smaller than under the any-relevant label, and each replacement has positive net gain. Figure~\ref{fig:strict} compares the two negative-flip rates on a shared vertical scale.

\begin{table}[h]
\caption{Strict HotpotQA transitions. Support requires every positive qrel passage in the top $k$.}
\label{tab:strict}
\centering
\small
\begin{tabular}{llrrrr}
\toprule
Model & $k$ & BM25 & Both & Fixed & Broken \\
\midrule
BGE & 1  & 0    & 0    & 0    & 0 \\
BGE & 5  & 1874 & 1543 & 2129 & 331 \\
BGE & 10 & 2432 & 2098 & 2180 & 334 \\
BGE & 20 & 2974 & 2640 & 2136 & 334 \\
BGE & 50 & 3644 & 3343 & 1978 & 301 \\
\midrule
E5 & 1  & 0    & 0    & 0    & 0 \\
E5 & 5  & 1874 & 1577 & 1964 & 297 \\
E5 & 10 & 2432 & 2134 & 2008 & 298 \\
E5 & 20 & 2974 & 2644 & 1943 & 330 \\
E5 & 50 & 3644 & 3325 & 1811 & 319 \\
\midrule
SPLADE & 1  & 0    & 0    & 0    & 0 \\
SPLADE & 5  & 1874 & 1468 & 1304 & 406 \\
SPLADE & 10 & 2432 & 2012 & 1338 & 420 \\
SPLADE & 20 & 2974 & 2511 & 1328 & 463 \\
SPLADE & 50 & 3644 & 3197 & 1193 & 447 \\
\bottomrule
\end{tabular}
\end{table}

\subsection{SPLADE Weight Cutoff}
\label{app:splade}

The main SPLADE results keep the 256 largest positive weights. This comparison uses one encoding of \texttt{naver/splade-cocondenser-ensembledistil}. Inputs are truncated to 256 tokens and scored by $\log(1+\mathrm{ReLU}(\cdot))$ with max pooling. One ranking keeps the 256 largest positive weights. The other keeps every positive weight. The two configurations differ by at most a small number of queries and yield the same qualitative conclusions: net gain remains positive and negative flips remain present across all evaluated settings. Table~\ref{tab:splade-cutoff} reports the transition counts from this encoding. The main tables retain the saved SPLADE rankings.

\begin{table}[h]
\caption{SPLADE transitions against BM25 when the 256-weight cutoff is removed. These counts are nearly identical to the main SPLADE results.}
\label{tab:splade-cutoff}
\centering
\small
\begin{tabular}{llrrrr}
\toprule
Corpus & $k$ & BM25 & Both & Fixed & Broken \\
\midrule
Natural Questions & 1  & 517  & 349  & 845  & 168 \\
Natural Questions & 5  & 1240 & 1118 & 1209 & 122 \\
Natural Questions & 10 & 1600 & 1511 & 1136 & 89 \\
Natural Questions & 20 & 1956 & 1897 & 1036 & 59 \\
Natural Questions & 50 & 2347 & 2309 & 845  & 38 \\
\midrule
HotpotQA & 1  & 4875 & 4458 & 1589 & 417 \\
HotpotQA & 5  & 6130 & 5969 & 945  & 161 \\
HotpotQA & 10 & 6484 & 6359 & 708  & 125 \\
HotpotQA & 20 & 6759 & 6674 & 509  & 85 \\
HotpotQA & 50 & 7025 & 6973 & 306  & 52 \\
\midrule
FiQA & 1  & 139 & 114 & 102 & 25 \\
FiQA & 5  & 246 & 219 & 130 & 27 \\
FiQA & 10 & 306 & 278 & 117 & 28 \\
FiQA & 20 & 365 & 332 & 110 & 33 \\
FiQA & 50 & 414 & 395 & 107 & 19 \\
\bottomrule
\end{tabular}
\end{table}

\end{document}